\documentclass[prl,twocolumn,superscriptaddress,floats]{revtex4}

\usepackage[dvips]{graphicx}
\usepackage{amsmath}
\usepackage{booktabs}
\usepackage{dcolumn}
\usepackage{sidecap}

\usepackage{rotating}

\newcolumntype{d}{D{.}{.}{-1}}

\newcommand{\grad}{\ensuremath{^\circ}}

\newcommand{\yca}{Y$_{1-x}$Ca$_{x}$TiO$_{3}$}

\newcommand{\rca}{$RE$$_{1-x}$Ca$_{x}$TiO$_{3}$}
\newcommand{\reca}{$RE$$_{1-x}$Ca$_{x}$TiO$_{3}$}

\begin{document}

\title{Evidence for charge and orbital order in the doped titanates \\ $RE$$_{1-x}$Ca$_x$TiO$_3$ ($RE$=Y, Er, Lu)}

\author{A. C. Komarek }
\author{M. Reuther }
\author{T. Lorenz}
\affiliation{II. Physikalisches Institut, Universit\"{a}t zu
K\"{o}ln, Z\"{u}lpicher Str. 77, D-50937 K\"{o}ln, Germany}
\author{A. Cousson }
\affiliation{Laboratoire L\'{e}on Brillouin, CEA/CNRS, F-91191 Gif
sur Yvette Cedex, France}
\author{P. Link}
\affiliation{Forschungsneutronenquelle Heinz Maier-Leibnitz
(FRM-II), TU M\"unchen,  D-85747 Garching, Germany}
\author{W. Morgenroth}
\affiliation{Geowissenschaften, Abteilung Kristallographie,
Universit\"at Frankfurt, Altenh\"oferallee 1, D-60438 Frankfurt,
Germany} \affiliation{Hasylab/DESY Notkestr. 85, D-22607, Hamburg,
Germany}
\author{D. Trots}
\author{C. Baehtz}\affiliation{Darmstadt Univ. of
Techn., Inst. f. Mat. Science, Petersenstr. 23, D-64287 Darmstadt,
Germany} \affiliation{Hasylab/DESY Notkestr. 85, D-22607 Hamburg,
Germany}
\author{M. Braden}
\affiliation{II. Physikalisches Institut, Universit\"{a}t zu
K\"{o}ln, Z\"{u}lpicher Str. 77, D-50937 K\"{o}ln, Germany}
\date{\today}

\pacs{PACS numbers:}

\begin{abstract}

Combining macroscopic and diffraction methods we have studied the
electric, magnetic and structural properties of
$RE$$_{1-x}$Ca$_x$TiO$_3$ ($RE$=Y, Er, Lu) focusing on the
concentration range near the metal-insulator transition. The
insulating phase, which is stabilized by a smaller rare-earth
ionic radius, exhibits charge order with a predominant occupation
of the $d_{xy}$ orbital. The charge and orbital ordering explains
the broad stability range of the insulating state in
$RE$$_{1-x}$Ca$_x$TiO$_3$ with smaller rare-earth ions. The strong
modulation of the Ti-O bond distances indicates sizeable
modulation of the electric charge.

\end{abstract}
\maketitle

The perovskite titanates $RE$TiO$_3$ ($RE$ a rare earth) can be
considered as the one-electron analogue to the parent compounds of
the cuprate superconductors with a single hole in the 3d shell
\cite{imada}, but the orbital degree of freedom is not quenched in
the titanates. The series $RE$TiO$_3$ exhibits a transition from
an antiferromagnetic Mott insulator for $RE$=La to a ferromagnetic
insulator for smaller $RE$ (for example YTiO$_3$) which is
associated with a change in the 3d orbital occupation
\cite{imada,2}. Hole doping suppresses the Mott-insulating
ground-states rendering the samples paramagnetic and metallic.
However, the phase diagrams of the two most studied systems,
La$_{1-x}$Sr$_{x}$TiO$_{3}$ and Y$_{1-x}$Ca$_{x}$TiO$_{3}$, differ
considerably\cite{imada}. Whereas a small amount of Sr
(approximately 5\%) is sufficient to render LaTiO$_3$ metallic,
\yca \ samples stay insulating up to Ca-doping levels of about
40\% (at room-temperature) \cite{imada,4,4b,5}. This astonishing
stability of the insulating state in the \yca \ series can only
approximately be attributed to reduced electronic band widths. The
smaller A-site ionic radius causes larger structural distortions
which reduce the electronic band width. However, this band-width
reduction seems to be insufficient to explain the remarkable
stability of the insulating state up to high amounts of doping in
Y$_{1-x}$Ca$_{x}$TiO$_{3}$. By using various diffraction methods
on single crystals of \rca , we elucidate the anomalous extension
of the insulating state which exhibits charge and orbital
ordering. This electronic order stabilizes the insulating phase
similar to the phase diagrams of Pr$_{1-x}$Ca$_x$MnO$_3$
\cite{manganate} and La$_{2-x}$Sr$_x$NiO$_4$ \cite{nickelate}.

\begin{figure}[h]
\begin{center}
\includegraphics*[width=0.95\columnwidth]{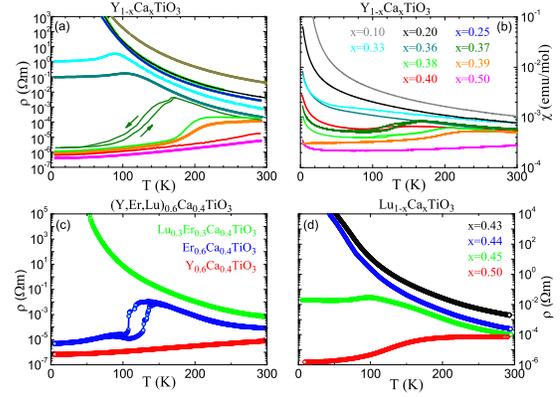}
\end{center}
{~~~~} \vskip -1 true cm\caption{(color online) The temperature
dependencies of the resistivity (a) and of the magnetic
susceptibility (b) in the \yca \ series, respectively. The
electric resistivity is shown for various
$RE$$_{0.6}$Ca$_{0.4}$TiO$_3$ samples in (c) and for
Lu$_{1-x}$Ca$_{x}$TiO$_3$ in (d).} \label{figpeaks}
\end{figure}

Extensive studies of the metal-insulator (MI) transition in \yca \
\cite{4,4b,5,6,7,8,9} revealed remarkable properties. Samples with
a Ca content at the border of metallic and insulating phases
exhibit a temperature-driven  MI transition from an insulating
phase at high temperature towards a metallic phase at low
temperature. This phase sequence is rather uncommon as most
materials exhibit the reversed sequence with the insulating phase
being stable at low temperatures. In addition, there is clear
evidence for phase segregation \cite{4,5} with coexistence of
metallic and insulating phases at low temperature.
The unknown cause of the enhanced stability
of the insulating state in \yca \ motivated us to perform
comprehensive diffraction studies.

Untwinned single crystals of \reca\ ($RE$=Y, Er, and Lu) have been
grown using a floating-zone image furnace. Special effort was
undertaken to control the stoichiometry which was verified by EDX
analyses, by thermogravimetry and by X-ray single-crystal
diffraction \cite{dr-roth,dr-komarek}. The oxygen content
determined by TGA corresponds to 3.000$\pm$0.005 per formula unit
for most samples studied in this work. The electrical resistivity
was measured with a four-probe technique and the magnetic
susceptibility by a vibrating sample magnetometer. X-ray powder
diffraction data were collected on a D5000 laboratory
diffractometer using Cu K$_\alpha$ radiation and on the beam-line
B2 at the Hasylab/DESY synchrotron (wavelength 0.7495\AA ).
Single-crystal experiments were performed with a Bruker Apex X8
diffractometer, with synchrotron radiation (beam-line D3 at the
Hasylab/DESY), and with the 5C2 neutron four-circle diffractometer
at the Laboratoire L\'eon Brillouin and with the PANDA
spectrometer at FRM-II.

Figure 1 shows the temperature dependencies of the electric
resistivity and of the magnetic susceptibility in \yca . For
x=0.10, 0.20, and 0.25 the samples exhibit the characteristic
insulating increase of electric resistivity upon cooling, but for
x=0.33 and 0.36 the resistivity saturates towards low
temperatures. Further increase of the Ca content leads to metallic
behavior at low temperature with a pronounced hysteresis. The MI
transition is also seen in the susceptibility which is much
smaller in the metallic phase. Finally samples with x=0.4 and 0.5
exhibit metallic resistivity over the entire temperature range
accompanied by a low magnetic susceptibility. These results agree
with earlier studies\cite{6,7,8,9}. The insulating phase is
further stabilized when Y is replaced by smaller rare-earth ions,
see Fig. 1(c) and 1(d). While Y$_{0.6}$Ca$_{0.4}$TiO$_3$ remains
metallic till room temperature, Er$_{0.6}$Ca$_{0.4}$TiO$_3$
exhibits the MI transition at about 100K, and finally
Lu$_{0.3}$Er$_{0.3}$Ca$_{0.4}$TiO$_3$ is insulating at all
temperatures. This effect is also seen in the
Lu$_{1-x}$Ca$_{x}$TiO$_3$ series, which exhibits the MI transition
at a higher Ca content compared to \yca , see Fig. 1(d).

\begin{figure}[h]
\begin{center}
\includegraphics*[width=0.92\columnwidth]{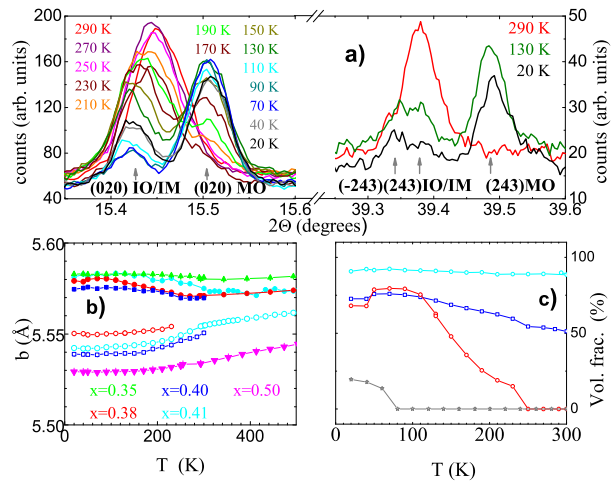}
\end{center}
{~~~~} \vskip -3.2 true cm \caption{(color online)
 (a) Diffraction patterns obtained with synchrotron radiation at
beam-line B2 at HASYLAB for the sample with x=0.38. On the left
the (020)-peaks of the insulating IO/IM phase and of the MO phase
are shown for different temperatures. On the right side the
($\pm$243) peaks are shown. As the peaks of the insulating and
metallic phases are well separated, one can easily observe a
splitting in the left peak indicating the transition from the IO
phase to the IM phase with $\beta\approx 90.1$\grad. (b) $b$
lattice parameter as a function of temperature. The samples with
x=0.35 and x=0.38 were measured at beam-line B2 at HASYLAB, others
with a laboratory source. Open symbols denote values for the MO
phase.  (c) The volume ratio of the MO phase for x=0.37 (taken
from referece \cite{4b} (black), 0.38 (red), 0.40 (blue), and 0.41
(light blue). } \label{figpeaks}
\end{figure}

Powder-diffraction data are shown in Fig. 2 for the composition
Y$_{0.62}$Ca$_{0.38}$TiO$_3$. At high temperature, the diffraction
patterns can be perfectly described by an orthorhombic lattice
(space group Pbnm), but peaks split at low temperature indicating
phase segregation. The additional phase can be described in space
group Pbnm down to the lowest temperatures, but it exhibits a
smaller $b$ unit-cell parameter, see Fig. 2(b), and a smaller
unit-cell volume (not shown). The original phase with the larger
unit-cell volume transforms into a monoclinic phase as can be seen
in the additional splitting of the corresponding peaks, see Fig.
2(a) and in the temperature dependence of the monoclinic angle
$\beta$. The phase segregation and the transition into a
monoclinic phase perfectly agree with a previous diffraction study
on \yca \cite{4,5}. The structural transition coincides with the
anomalies seen in the electric resistivity and in the magnetic
susceptibility. With the temperature dependence of the volume
fraction of the two phases, see Fig. 2(c), we may simultaneously
describe the magnetic and resistivity data. Thereby, one can
unambiguously assign the small unit-cell volume phase to the
metallic phase which we label MO (metallic orthorhombic) in the
following. The insulating phase thus exhibits the additional
structural phase transition from the IO (insulating orthorhombic)
phase at high temperature to the IM (insulating monoclinic) phase
at low temperatures.

The microscopic character of the monoclinic phase has been
elaborated by a series of diffraction experiments using neutron,
synchrotron and laboratory X-ray radiation. Due to the $b$
glide-mirror plane in space group Pbnm, (0kl) reflections with $k$
odd are extinct. Neutron diffraction experiments on the PANDA
spectrometer, however, show these reflections to appear at a
temperature of about 150\ K in \yca \ with x=0.33, 0.35, and 0.36,
see Fig. 3. The (011) and (013) reflections can be transformed
into cubic notation referring to the simple 3.8\AA \ cubic
perovskite cell, where they correspond to (0.5,0.5,0.5)$_{cubic}$
and (0.5,0.5,1.5)$_{cubic}$, respectively. These half-integer
indexed reflections refer to a G-type distortion, i.e. a
three-dimensional checkerboard ordering, which occurs in the
monoclinic phase.

\begin{figure}[t]
\begin{center}
\includegraphics*[width=0.85\columnwidth]{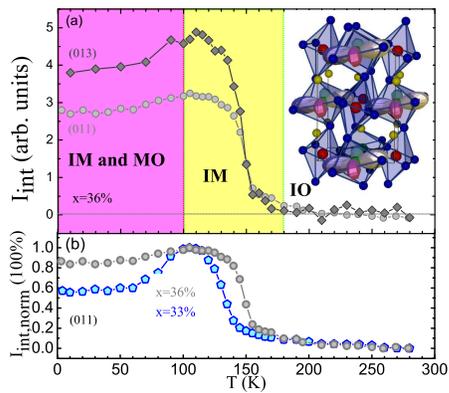}
\end{center}
\caption{(color online) (a) Intensities of the (011) and (013)
superstructure reflections in Pbnm for
Y$_{0.64}$Ca$_{0.36}$TiO$_3$ measured on the PANDA neutron
spectrometer.  The inset shows the $\sqrt(2)$x$\sqrt(2)$x2 unit
cell of the charge-ordered structure with Ti$^{4+}$ ions
\emph{(red)} and Ti$^{3+}$ ions \emph{(green)} and the
predominantly occupied $d_{xy}$ orbital at the Ti$^{3+}$ site. (b)
shows the comparison of the (011) intensity measured on PANDA for
x=0.33 and x=0.36.} \label{figPanda}
\end{figure}

Full structure analyzes have been performed by single-crystal
diffraction, see table I. The  \yca \ crystals with x=0.35 and
0.36 have been studied at room temperature in the IO phase and at
low temperature in the IM phase collecting large
Bragg-reflection-intensity data sets on the neutron four-circle
diffractometer 5C2. The refinements of the crystal structure
models in space groups Pbnm and P2$_1$/n yield no significant
difference for the room-temperature data sets, but the
low-temperature data is much better described in space group
P2$_1$/n. In this lower symmetry, there are two inequivalent Ti
sites, labelled Ti1 and Ti2, and two in-plane oxygen sites,
labelled O2 and O2'. The structure refinements yield a clear
difference in the oxygen surrounding with a larger mean Ti-O
distance for the Ti2 site. Whereas the Ti1 site shows an almost
isotropic bonding, the octahedron around Ti2 is flattened due to
stretching of the bonds parallel to the $a,b$ plane. These oxygen
coordinations can be further analyzed by the bond-valence sum
(BVS) formalism \cite{bvs}, where the empirical valence per bond
is summed over the coordination,
$Z_{BVS}=\sum_{i}{e^{{r_0-r_i\over B}}}$ with $r_i$ the bond
distance, $r_0$ and $B$ empirical parameters. The BVS analysis
indicates a modulation of the charge at the Ti sites by
$\Delta$Z=0.32(1) and 0.23(1) electron charges, for x=0.35 and
0.36 respectively. In addition the flattened shape of the TiO$_6$
octahedron around Ti2 points to an orbital ordering with a
predominant occupation of the $d_{xy}$ orbital. These structural
distortions show that \yca \ in the IM phase exhibits charge
ordering with additional orbital polarization occurring at the
lower-valent site Ti2 (nominally $3+$). The distorted crystal
structure is drawn in Fig. 3 (a).

\begin{figure}[t]
\begin{center}
\includegraphics*[width=0.72\columnwidth]{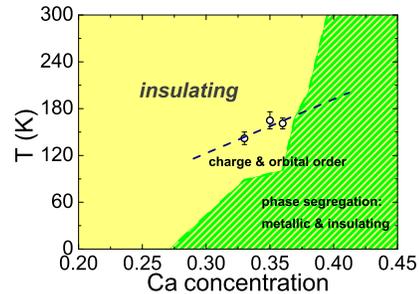}
\end{center}
\caption{(color online) Phase diagram of \yca . MI transition
temperatures were obtained from the resistivity data, and
transition temperatures for charge and orbital ordering were
determined in the neutron diffraction experiments. \yca \ exhibits
coexistence of the metallic and the insulating phase in a broad
concentration range and only highly-doped samples show an almost
single metallic phase. } \label{figN}
\end{figure}

Figure 4 presents the phase diagram of \yca \ near the MI
transition. The transformation into the metallic state upon
cooling is not complete but results in phase segregation and in
phase coexistence. The insulating phase exhibits the additional
transition into the low-temperature monoclinic phase driven by
charge and orbital ordering. The charge modulation in \yca \ as
obtained by the BVS calculations is smaller than one electron
charge, which may not be expected as the electronic doping of
these compounds is not sufficient for integer-valent checkerboard
ordering. Increase of the doping stabilizes the metallic phase in
\yca \ and no trace of the insulating monoclinic and
charge-ordered phase can be found for Y$_{0.5}$Ca$_{0.5}$TiO$_3$
by diffraction techniques. However, the charge ordering at an
electronic doping close to 50\% can be studied in the compounds
containing small rare-earth ions, in which the insulating phase is
further stabilized, see Fig. 1(c) and 1(d). X-ray single-crystal
diffraction experiments have been performed for
Lu$_{0.3}$Er$_{0.3}$Ca$_{0.4}$TiO$_3$,
Lu$_{0.56}$Ca$_{0.44}$TiO$_3$, and  Lu$_{0.5}$Ca$_{0.5}$TiO$_3$ at
temperatures of 8K, 125K and 10 K, respectively. For the first two
data sets, the refinements significantly improve in the monoclinic
space group, whereas the Lu$_{0.5}$Ca$_{0.5}$TiO$_3$ data are
equally well described in space group Pbnm. This indicates that
Lu$_{0.3}$Er$_{0.3}$Ca$_{0.4}$TiO$_3$ and
Lu$_{0.56}$Ca$_{0.44}$TiO$_3$ exhibit charge and orbital ordering,
but Lu$_{0.5}$Ca$_{0.5}$TiO$_3$ not, in perfect agreement with the
respective behavior found in the electric resistivity, see Fig.
1(d), which shows that Lu$_{0.5}$Ca$_{0.5}$TiO$_3$ stays metallic
to the lowest temperatures. Qualitatively, the structural
distortions in Lu$_{0.3}$Er$_{0.3}$Ca$_{0.4}$TiO$_3$ and in
Lu$_{0.56}$Ca$_{0.44}$TiO$_3$ agree with those in the charge
ordered \yca \ compounds described above. In particular, we find
the identical flattening of the octahedron around the Ti2 site
which points to orbital order associated with predominant
occupation of the $d_{xy}$ orbital. The modulation of the
electronic charge is, however, significantly enhanced yielding
values of $\Delta$Z=0.72(3) and 0.82(2) electron charges for
Lu$_{0.3}$Er$_{0.3}$Ca$_{0.4}$TiO$_3$ and
Lu$_{0.56}$Ca$_{0.44}$TiO$_3$, respectively. This enhanced
electronic modulation can be attributed to the doping level close
to x=0.5 which allows for an ideal checker-board charge order with
$\Delta$Z=1.

\begin{table}[!b]\centering{
{\scriptsize
\begin{tabular}[t]{c|cccccc}
\hline\hline
 x & 0.35 & 0.35  & 0.36      & 0.36    & 0.4 &  0.44   \\
 T(K) & 120 & 298 &  110 &  298 &  8 &  125   \\
 radiation & n & n & n  & n  & synchr.  & X-ray   \\
 space group & P2$_1$/n & Pbnm &  P2$_1$/n  &  Pbnm &   P2$_1$/n &   P2$_1$/n  \\

 reflections & 1538 &1553 & 1790  & 366   & 17368 & 7853   \\
R/R$_w$ (\%)  &  4.0/8.4      & 3.5/5.8       &  3.3/3.1 & 5.4/5.9
& 2.8/5.5       &  2.3/4.3 \\
\hline $a$ (\AA)        &   {\tiny   5.3579(1) }      &    {\tiny
5.3679(1) }      & {\tiny  5.3461(1)}       &   {\tiny  5.3553(1)}
&     {\tiny  5.3317(5)}
 & {\tiny  5.3123(2)}    \\
$b$ (\AA)        &   {\tiny   5.5804(1) }     &    {\tiny
5.5834(1) }      & {\tiny  5.5865(1)}     &    {\tiny  5.5823(1)}
&     {\tiny  5.5667(5)}
& {\tiny  5.5542(2)}    \\
$c$ (\AA)        &   {\tiny   7.6629(1)}       &   {\tiny
7.6804(1) }      & {\tiny  7.6402(1)}       &   {\tiny  7.6579(1)}
&     {\tiny  7.6452(5)}
 & {\tiny  7.6273(3)}    \\
\hline x(Y1/Ca1)        &{\tiny  0.9834(1)}&{\tiny
0.9841(1)}&{\tiny  0.9847(1)}&{\tiny
0.9849(2)}& {\tiny  0.9814(1)}   & {\tiny  0.9812(1)}  \\
y(Y1/Ca1)        &{\tiny  0.0641(1)}&{\tiny  0.0631(1)}&{\tiny
0.0623(1)}&{\tiny
0.0612(4)}& {\tiny  0.0661(1)} & {\tiny  0.0660(1)}  \\
z(Y1/Ca1)        &{\tiny  0.2476(2)}&{\tiny  0.25     }&{\tiny
0.2488(3)}&{\tiny
0.25     }& {\tiny  0.2506(1)}  & {\tiny  0.2504(1)}  \\
U(Y1/Ca1)        &{\tiny  0.0095(2)}&{\tiny  0.0136(2)}&{\tiny
0.0047(1)}&{\tiny
0.0063(7)}&  {\tiny  0.0041(1)} &  {\tiny  0.0022(1)}  \\
U(Ti1/Ti2)       &{\tiny  0.0085(3)}&{\tiny  0.0111(3)}&{\tiny
0.0029(1)}&{\tiny
0.0038(11)}&  {\tiny  0.0036(1)} &  {\tiny  0.0009(1)}  \\
x(O1)            &{\tiny  0.1039(1)}&{\tiny  0.1030(2)}&{\tiny
0.0999(1)}&{\tiny
0.1002(2)}&  {\tiny  0.1057(3)} &  {\tiny  0.1055(3)}  \\
y(O1)            &{\tiny  0.4656(1)}&{\tiny  0.4658(1)}&{\tiny
0.4678(1)}&{\tiny
0.4668(5)}&  {\tiny  0.4639(2)} &   {\tiny  0.4633(2)} \\
z(O1)            &{\tiny  0.2507(3)}&{\tiny  0.25     }&{\tiny
0.2526(4)}&{\tiny
0.25     }&  {\tiny  0.2499(6)} &  {\tiny  0.2471(5)} \\
U(O1)            &{\tiny  0.0125(2)}&{\tiny  0.0154(2)}&{\tiny
0.0068(1)}&{\tiny
0.0088(8)}&  {\tiny  0.0093(3)} &  {\tiny  0.0082(3)}  \\
x(O2)            &{\tiny  0.6993(3)}&{\tiny  0.6956(1)}&{\tiny
0.7004(3)}&{\tiny
0.6964(2)}&  {\tiny  0.7049(8)} &  {\tiny  0.699(1)}  \\
y(O2)            &{\tiny  0.3057(2)}&{\tiny  0.3013(1)}&{\tiny
0.3035(3)}&{\tiny
0.3003(3)}&  {\tiny  0.3068(7)} &  {\tiny  0.3108(5)}  \\
z(O2)            &{\tiny  0.0554(2)}&{\tiny  0.0529(1)}&{\tiny
0.0493(3)}&{\tiny
0.0517(1)}&  {\tiny  0.0553(6)} &   {\tiny  0.0491(4)} \\
U(O2)            &{\tiny  0.0125(2)}&{\tiny  0.0159(1)}&{\tiny
0.0066(1)}&{\tiny
0.0085(6)}&  {\tiny  0.0078(3)}  &  {\tiny  0.0065(3)}  \\
x(O3)            &{\tiny  0.8079(3)}&{\tiny     /     }&{\tiny
0.8077(3)}&{\tiny
 /     }&  {\tiny  0.8155(8)} &  {\tiny  0.809(1)}  \\
y(O3)            &{\tiny  0.7977(3)}&{\tiny     /     }&{\tiny
0.7965(3)}&{\tiny
 /     }&  {\tiny  0.7965(7)} &  {\tiny  0.7919(5)} \\
z(O3)            &{\tiny  0.0514(2)}&{\tiny     /     }&{\tiny
0.0540(3)}&{\tiny
 /     }&  {\tiny  0.0543(6)} & {\tiny  0.0609(4)}  \\
U(O3)            &{\tiny  0.0122(1)}&{\tiny     /     }&{\tiny
0.0066(1)}&{\tiny
 /     }&  {\tiny  0.0041(1)} & {\tiny  0.0065(3)}  \\
\hline
 Ti1-O1 (\AA )& 2.011(2)  & 1.993(1)& 2.018(3)  & 1.995(1)  & 2.002(4)  & 1.976(4) \\
 Ti1-O2 (\AA )&  1.991(1) &2.019(1) & 1.984(2)  & 2.004(1)  & 1.952(4)  & 1.950(4) \\
 Ti1-O2' (\AA )& 1.997(1) & 2.045(1)& 1.996(2)  & 2.005(1)  & 1.966(4)  & 1.967(3) \\
 Ti2-O1 (\AA )& 2.001(2)  &    /    & 1.980(3)  &     /     & 2.003(4)  & 2.020(4) \\
 Ti2-O2 (\AA )& 2.060(1)  &    /    & 2.044(2)  &     /     & 2.071(4)  & 2.059(3) \\
 Ti2-O2' (\AA )& 2.041(2) &    /    & 2.047(2)  &     /     & 2.070(4)  & 2.062(4) \\
 $\Delta$ BVS  & 0.32(1)   &    /    & 0.23(1)   &     /     & 0.72(3)  &  0.82(2) \\
\hline
  \end{tabular}
  } \\ }
 \caption{\label{tabA} Crystal structure resulting from various
 single-crystal diffraction experiments using neutron (n),
 synchrotron (synchr.) or laboratory X-ray radiation on \yca \
 (x=0.35 and 0.36), Lu$_{0.3}$Er$_{0.3}$Ca$_{0.4}$TiO$_3$ (x=0.4)
 and Lu$_{0.56}$Ca$_{0.44}$TiO$_3$ (x=0.44).}
\end{table}

The charge and orbital ordering in $RE$$_{1-x}$Ca$_x$TiO$_3$
strongly resembles the corresponding effects in several
manganites. Also in La$_{1-x}$Sr$_{1+x}$MnO$_4$ \cite{13} and in
perovskite manganites \cite{zimmermann,chen} commensurate charge
and orbital ordering of the checkerboard type occurs even below
the x=0.5 doping level. The charge and orbital ordered state
becomes more stable close to the optimum doping in the manganates
\cite{13,senff2008} as well as in the titanates, but smaller $RE$
ions are needed in order to overcome the trend towards metallicity
in the titanates. There is also strong resemblence with the
$RE$NiO$_3$ series which exhibit a MI transition into a
charge-ordered phase except for $RE$=La \cite{14}. The charge
disproportionation in $RE$NiO$_3$ \cite{15,15a} is qualitatively
comparable to the charge order in $RE$$_{1-x}$Ca$_x$TiO$_3$, but
the titanates exhibit a stronger charge modulation close to the
ideal half-doping. Also in the nickelate series the $RE$ ionic
radius has a strong impact on the charge ordering, as the
distortions related with a smaller rare earth stabilize charge
order similar to our observations for $RE$$_{1-x}$Ca$_x$TiO$_3$.

So far no evidence for charge ordering has been reported for any
titanate bulk materials with the perovskite or with a related
layered structure. However, charge order is discussed for the
heterostructures involving LaTiO$_3$ and SrTiO$_3$ \cite{16,17}.
At the interface between these two materials, the same electronic
doping as for half-doped $RE$$_{1-x}$Ca$_x$TiO$_3$ with x=0.5 is
realized. This interface has been studied by density-functional
theory (DFT) \cite{17} calculations which obtain a checkerboard
ordering similar to our finding for the bulk material. Also the
orbital ordering obtained in the DFT interface calculations
perfectly agrees with our structure analyses, as a predominant
occupation of the $d_{xy}$ orbitals is found in both cases. The
observation of charge and orbital order in bulk titanates near
half doping gives strong support to the DFT interface calculations
\cite{17} and renders similar calculations for the bulk materials
highly desirable. The electron-phonon coupling associated with the
charge order in the titanates must be strong and certainly is a
relevant factor also in the physics of the interfaces.

In conclusion,  comprehensive diffraction studies on single
crystals of $RE$$_{1-x}$Ca$_x$TiO$_3$ ($RE$=Y, Er, Lu) identify
the origin of the monoclinic phase appearing at the
insulator-to-metal crossover. Charge and orbital ordering occurs
in samples with a smaller rare-earth ion and a doping level near
x=0.5. Upon cooling, the compositions close to the doping-driven
MI transition exhibit phase segregation into a metallic and an
insulating phase similar to manganates exhibiting colossal
magneto-resistivity. This insulating phase exhibits G-type charge
order. We may thus identify charge and orbital ordering as the
important element to stabilize the insulating state in the doped
titanates. Charge and orbital order should be a more general
phenomenon in titanate materials also relevant for the discussion
of heterostructures involving LaTiO$_3$ and SrTiO$_3$.

\par
This work was supported by the Deutsche Forschungsgemeinschaft
through SFB 608. We thank H. Roth for providing three of the
samples studied.

\end{document}